# An overdetermined linear equations-based time calibration method for fast sampling ASICs

Boyu Cheng [a,b], Lei Zhao [a,b]*, Jiajun Qin [a,b], Han Chen [a,b], Yuxiang Guo [a,b], Shubin Liu [a,b], Qi An [a,b]

[a] State Key Laboratory of Particle Detection and Electronics, University of Science and Technology of China, Hefei 230026, China
[b] Department of Modern Physics, University of Science and Technology of China, Hefei 230026, China

*Corresponding author. Tel: +86 0551 63607746. *E-mail address*: zlei@ustc.edu.cn (L. Zhao).

*Abstract*—A novel time calibration method for waveform sampling application specific integrated circuits (ASICs) based on switched capacitor arrays (SCAs) is proposed in this paper. Precision timing extraction using SCA ASICs has been proved to be a promising technique in many high energy physics experiments. However, such ASICs suffer from irregular sampling intervals caused by process variations. Therefore, careful calibration is required to improve the time resolution of such systems. We evaluated the limitation of a popular method using the proportionality between voltage amplitude and sampling interval of adjacent switched-capacitor cells responding to either a sawtooth wave or a sine wave. The new time calibration method presented here utilizes the relationship between sampling intervals and the known input signal period to establish overdetermined linear equations, and the roots of these equations correspond to the actual sampling intervals. We applied this new method to a pulse timing test with an ASIC designed by our team, and the test results indicate that the new time calibration method is effective.

## I. Introduction

High-precision time measurements are widely required in particle physics, such as time-of-flight (TOF) systems [1] [2] [3], neutrino physics [4], and gamma-ray astronomy [5]. They also play important roles in positron emission tomography (PET) systems [6]. Conventional techniques dedicated to precise time measurements are mainly based on the use of constant fraction discriminators (CFDs) associated with time-to-digital converters (TDCs). However, in recent years, some studies show that waveform sampling offers a better time resolution [3] [7]. Traditionally, fast analog-to-digital converters (ADCs) are used for waveform digitization. As an alternative method, switched capacitor array (SCA) ASICs are becoming increasingly popular as a front-end readout solution for many high energy physics experiments [8] [9] [10] [11] [12], because of their high sampling speed, high channel density, low power consumption, and affordable cost. As the sampling rates of such ASICs have been increased to several gigasamples per second (Gsps) [10] [12] or even more than 15 Gsps [13], methods have been developed based on waveform digitization with SCA ASICs in precise time measurements, and such methods provide a time resolution far better than the sampling frequency [13][14][15].

The T0 detector is one of the key components in the external target experiment of the cooling storage ring (CSR) in the heavy ion research facility in Lanzhou (HIRFL). Furthermore, a high precision time measurement is required in the readout electronics of T0 detector with a time precision of better than 25 ps RMS. The waveform digitization method based on SCA architecture is a promising solution. We designed a fast sampling ASIC with a sampling speed of up to 5.2 Gsps for this detector.

In Gsps SCA chips, the sampling clock pulses are generated in most cases by an inverter delay chain, including our own ASIC. The sampling interval between two adjacent cells is therefore determined by the transition time of the inverters between them. However, the inverters have irregular but fixed transition times due to process variations. The nonuniform sampling intervals deteriorate the resolution of time measurements; therefore, a proper time calibration is required. Several time calibration methods proposed for SCA ASICs are reported in Refs. [13] [16] [17] [18] [19] [20] . The basic ideas in Refs. [13] [16] [19] [20] are similar; they all use the proportionality between voltage amplitude and sampling interval of adjacent switched-capacitor cells responding to sine waves or sawtooth waves. Furthermore, Ref. [18] is also based on this idea, and further improves the time resolution by applying a global time calibration.

These methods are proved to be effective; however, there are still some limitations. The analysis is presented in Section II. Based on the analysis, we propose a new time calibration method. In Section III, an SCA ASIC is introduced and setup of the test bench is described. In Section IV, the time resolution test results are shown. Finally, the conclusion is provided in Section V.

## II. Time Calibration Method

In this section, we first analyze the limitation of a typical time calibration method, and then propose our new method. The simulation results of these two methods based on MATLAB are also described and a comparison between them is made.

## A. Analysis of the typical time calibration method

Detailed discussions about the typical calibration method are reported in Refs. [16] [18] [20]. The principle of this method is to sample a sine wave (or a sawtooth wave) and measure the amplitude difference between two adjacent samples surrounding a zero-crossing point of the digitized waveform, and we address this method as the zero-crossing method in the following discussion. Because this amplitude difference is directly proportional to the sampling interval between the two cells, the sampling intervals can be calibrated. The basic idea is briefly described as follows and shown in Fig. 1.

Assuming that the sine input wave can be expressed as

$$V(t) = V_0 \cdot \sin(\omega t + \varphi) \quad (1),$$

where $V_0$, $\omega$, and $\varphi$ are the amplitude, angular frequency, and phase, respectively. Then, we can obtain the relationship between the sampling interval and the amplitude difference, as in

$$p = \frac{\Delta V_1}{\Delta t_1} = \frac{\Delta V_2}{\Delta t_2} = \cdots = \frac{\Delta V_n}{\Delta t_n}, n = 1, 2 \ldots N \quad (2),$$

$$\sum_{n=1}^{N} \Delta t_n = T_{clc} \quad (3), \text{ and}$$

$$\Delta t_n = \frac{\Delta V_n}{\sum_{j=1}^{N} \Delta V_j} \cdot T_{clc} \quad (4),$$

where $N$ is the cell number of one SCA channel, $\Delta V_n$ is the amplitude difference between the two adjacent samples surrounding the "zero-crossing" at cell $n$ and cell $n+1$, $\Delta t_n$ is the sampling interval between them, $T_{clc}$ is the reference clock period of the delay-locked loop (DLL), and $p$ is the slope of the sine signal at the "zero-crossing" point.

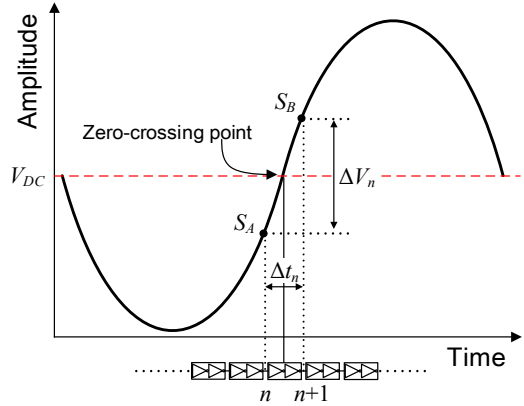

Fig. 1. Waveform sampling using an SCA.

However, in our research, we found that the hypothesis shown in Eq. (1) is not strictly correct. As for SCA ASICs, the input signal is transmitted along a path acting like a signal bus that all the sampling cells are connected to, and therefore parasitic resistance and capacitance are inevitable, as shown in Fig. 2. The structure in Fig. 2 actually forms low pass filters and causes an amplitude attenuation along the signal bus. In addition, the process variations also contribute to the gain fluctuation. Therefore, the waveform observed by the cells inside the SCA is

$$V_n(t) = G_n V_0 \cdot \sin(\omega t + \varphi) \quad (5)$$

where $G_n$ is the voltage gain of cell $n$. Now the slope $p_n$ at the "zero-crossing" point is.

$$p_n = G_n \cdot p \quad (6)$$

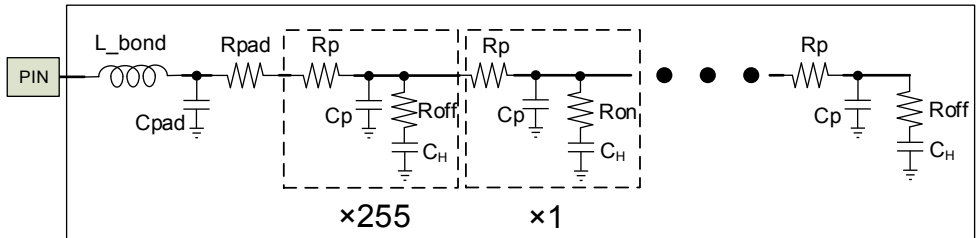

Fig. 2. Circuit model of parasitic impedances of the SCA structure, where $R_p$ is the parasitic resistance, $C_p$ is the parasitic capacitance, $R_{on}$ is the on-resistance of sampling switch, $R_{off}$ is the off-resistance, and $C_H$ is the sampling capacitor.

In Ref. [18], the authors proposed a "global" calibration method to enhance the calibration precision of the sampling intervals by comparing the time interval (obtained through calibration) between two adjacent zero-crossing points with the known input sine wave period. The time difference between them is used to correct the sampling interval mismatch among the cells. This procedure can be repeated iteratively many times until a good precision result is obtained.

*B. New time calibration method*

Inspired by the "global" calibration method, we propose a new method in this paper. In the first step, we calibrate the voltage mismatch errors among the sampling cells, which is the same as the traditional methods. Fig. 3 shows a sampled 202 MHz sine wave (after the voltage calibration) with a DC voltage $V_{DC}$ of 500 mV. Then we analyze the digitized waveform and calculate the sampling time interval. Points 'a' and 'b' are two adjacent zero-crossing points on the rising edge of the digitized sine wave signal. The time interval between 'a' and 'b' $\Delta t_{ab}$ can be expressed as

$$\Delta t_{ab} = \alpha \Delta t_{172} + \sum_{i=173}^{197} \Delta t_i + \beta \Delta t_{198} \tag{7}$$

where $\Delta t_i$ is the sampling interval between cell $i$ and cell $i+1$, $\alpha$ and $\beta$ are two voltage ratios given by a linear interpolation, as in

$$\alpha = \frac{V_{173} - V_{DC}}{\Delta V_{172}}, \beta = \frac{V_{DC} - V_{198}}{\Delta V_{198}} \tag{8}$$

where $V_i$ is the signal voltage value sampled at cell $i$, and $\Delta V_i$ is the amplitude difference between cell $i$ and cell $i+1$. The two ratios are almost unaffected by the gain variation since the two cells are adjacent to each other.

In fact, the time interval between 'a' and 'b' $\Delta t_{ab}$ should be equal to the period of input sine wave signal (marked as $T_C$), and we can obtain the following equation:

$$\alpha \Delta t_{172} + \sum_{i=173}^{197} \Delta t_i + \beta \Delta t_{198} = T_C \tag{9}$$

By repeating this procedure for a certain amount of times, we can obtain the overdetermined linear equations expressed as a matrix equation as shown in Eq. (10), where $M$ is the number of equations, $N$ is the aforementioned cell number of one SCA channel, and $M > N$. For overdetermined linear equations $Ax = b$, where $A$ is an equation coefficient matrix, $x$ and $b$ are vectors, we can use $A^T A x = A^T b$ to transform the overdetermined linear equations to regular linear equations, and then the only roots can be obtained. These roots ($\Delta t_1, \Delta t_2, \ldots \Delta t_N$) are the least squares solutions of the overdetermined linear equations. Compared with the global calibration method mentioned in Ref. [18], we do not need to obtain the "local" time calibration result as the first step, and thus boost the calibration efficiency and accuracy considerably without loop iterations.

$$\begin{pmatrix} 0 & \cdots & 0 & \alpha_1 & 1 & \cdots & 1 & \beta_1 & 0 & \cdots & 0 \\ 0 & \alpha_2 & 1 & \cdots & 1 & \beta_2 & 0 & 0 & 0 & \cdots & 0 \\ \cdots & 1 & \beta_3 & 0 & 0 & \cdots & 0 & 0 & \alpha_3 & 1 & \cdots \\ & & & & & \vdots & & & & & \\ 1 & \cdots & 1 & \beta_M & 0 & 0 & \cdots & 0 & \cdots & 0 & \alpha_M \end{pmatrix} \cdot \begin{pmatrix} \Delta t_1 \\ \Delta t_2 \\ \Delta t_3 \\ \vdots \\ \Delta t_N \end{pmatrix} = \begin{pmatrix} T_C \\ T_C \\ T_C \\ \vdots \\ T_C \end{pmatrix} \tag{10}$$

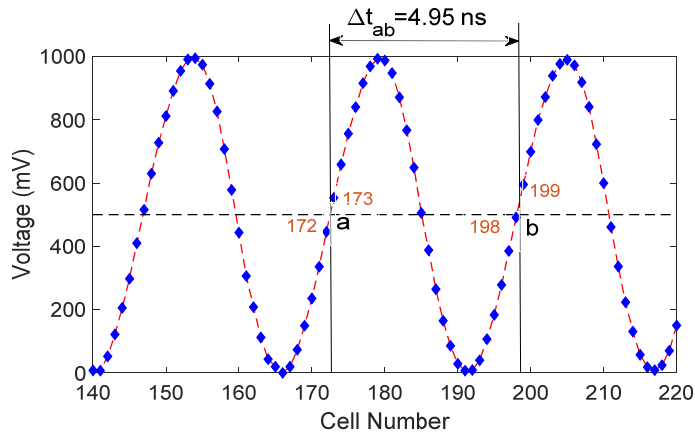

Fig. 3. Digitized waveform of a sine wave signal at a sampling speed of 5.2 Gsps.

## C. Simulations

To estimate the effect of our calibration method, a simulation test bench based on MATLAB was established. The gain variation of each cell caused by process and bandwidth limitation corresponds to nonideal parameters in the simulation, and the noise is not considered (does not need to be considered for calibration). The simulation data were generated with a sampling speed of 5 Gsps, i.e. the sampling intervals are 200 ps. The simulated input signal is a 202 MHz sine wave with an amplitude of 500 mV and a DC value of 500 mV. In our simulation, the circuits model shown in Fig. 2 was used to estimate the amplitude attenuation inside the chip with the parameters of $R_p = 0.3\ \Omega$, $C_p = 30\ fF$, $R_{on} = 250\ \Omega$, $C_H = 200\ fF$, and $R_{off} = 1\ M\Omega$. The values of these parameters come from the actual SCA ASIC that our team had designed, which will be briefly introduced later. Besides, the gain variations of the sampling cells are set by considering the process variations, which are uniformly distributed between 0.99 and 1.01. A total of 2000 waveforms were generated using MATLAB. Then, both the zero-crossing method and our new method were used to calibrate and correct the sampling intervals. The simulation results are shown in Fig. 4.

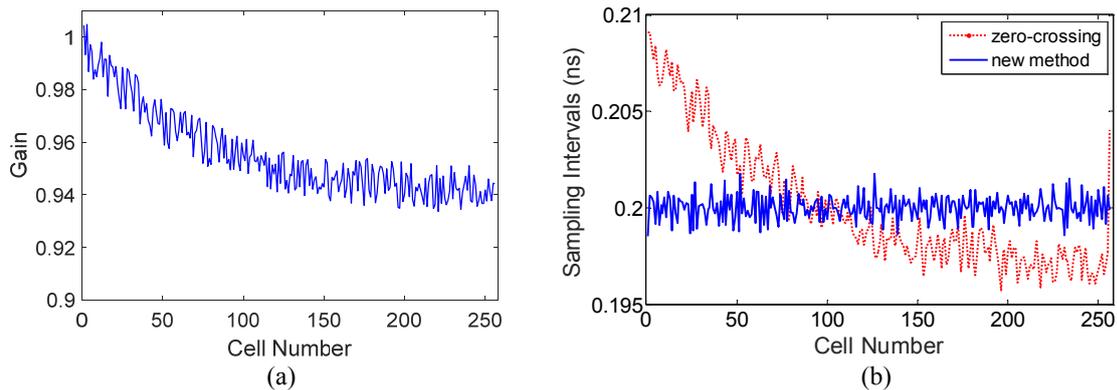

Fig. 4. Simulation results. (a) Voltage gain of each cell. (b) Sampling intervals of 256 cells after correction using the zero-crossing method and our new method.

Fig. 4 (a) shows the voltage gain of each cell used for simulation, and the sampling intervals calculated with the zero-crossing method in Fig. 4 (b) show a trend the same as the voltage gain, which is consistent with Eq. (4), and deviate from the ideal values. A large time-step at the last cell is caused by the large gain difference between the last cell and first cell. As shown in Fig. 4 (b), after calibration and correction using our method, the sampling intervals are quite close to the ideal 200 ps, which indicates good effect of this new method.

## III. TEST BENCH SETUP

We also conducted tests on our calibration method. We applied this calibration procedure on the SCA chip that our team had designed, the block diagram of which is shown in Fig. 5 (a). This SCA ASIC, designed in a 0.18 *μm* CMOS process, integrates eight channels with 256 sampling cells in each. This chip features a sampling speed of up to 5.2 Gsps, with sampling clock pulses generated by an on-chip delay-locked loop (DLL). The sampled voltages on the capacitors are digitized in parallel using the Wilkinson ADCs inside the chip, where the stored voltages are compared with a voltage ramp using the in-cell comparators and converted into output pulses from the comparators. The time differences between the start of ramp and the leading edges of these pulses are measured using a high-speed 12-bit gray counter. The results are stored in the latches within the cells, and further read out under the control of a "token" signal.

Fig. 5 (b) shows the test bench. A customized ASIC test board was designed, and a Xilinx Spartan-6 field programmable gate array (FPGA) based evaluation board was used for chip configuration and data processing. The trigger signals were generated by utilizing the discriminators on board and with their thresholds set by a 14-bit DAC. These trigger signals were used to control each channel individually after an appropriate delay configured by the FPGA. The input signals were generated by a Keysight 81160A waveform generator. In the tests, the SCA ASIC worked at a sampling speed of 5.2 Gsps.

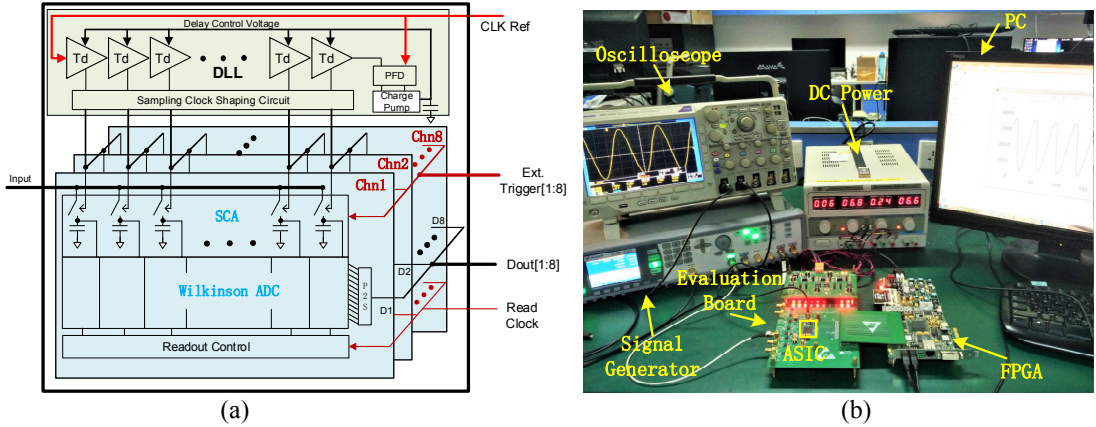

(a)                                            (b)

Fig. 5. (a) Block diagram of the SCA. (b) Photograph of the test bench.

## IV. TEST RESULTS

Voltage calibrations are required before time calibration. This process is in a traditional way, and therefore will not be presented in detail. In this section, we present the results of time calibration and performance tests.

### A. Sampling interval calibration results

As mentioned in Section II.A, there exists an amplitude attenuation along the signal bus, and we conducted tests to verify our analysis. We used the signal source to generate sine wave signals of 11MHz, 101MHz, and 202 MHz as the inputs for the SCA ASIC. A total of 500 digitized waveforms were overlaid together as shown in Fig. 6 (a)-(c), and a significant amplitude attenuation can be observed along the signal bus. Besides, as the input signal frequency increases, the amplitude attenuation becomes more significant. This is a frequency-dependent phenomenon that cannot be directly corrected for a pulse input which contains energy over a wide frequency band.

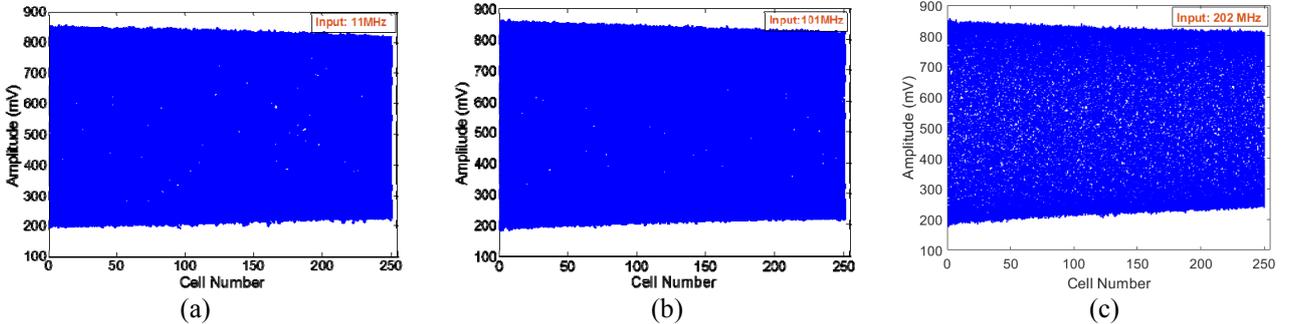

(a)                              (b)                            (c)

Fig. 6. Overlaid sine waves at 256 sampling cells. (a) With 11 MHz input signals. (b) With 101 MHz input signals. (c) With 202 MHz input signals.

Then we applied our new method proposed in Section II.B for time calibration. We used a total of 2000 digitized sine wave waveforms (202MHz) to establish the overdetermined linear equations and further calculated the sampling intervals. The calibration results are shown in Fig. 7 (a), which indicates an averaged sampling interval of 192.0 ps. Fig. 7 (b) shows its integrated nonlinearity (INL) and differential nonlinearity (DNL). The INL varies from -244 ps to 75 ps, and the DNL is between -67.8 ps and 93.1 ps.

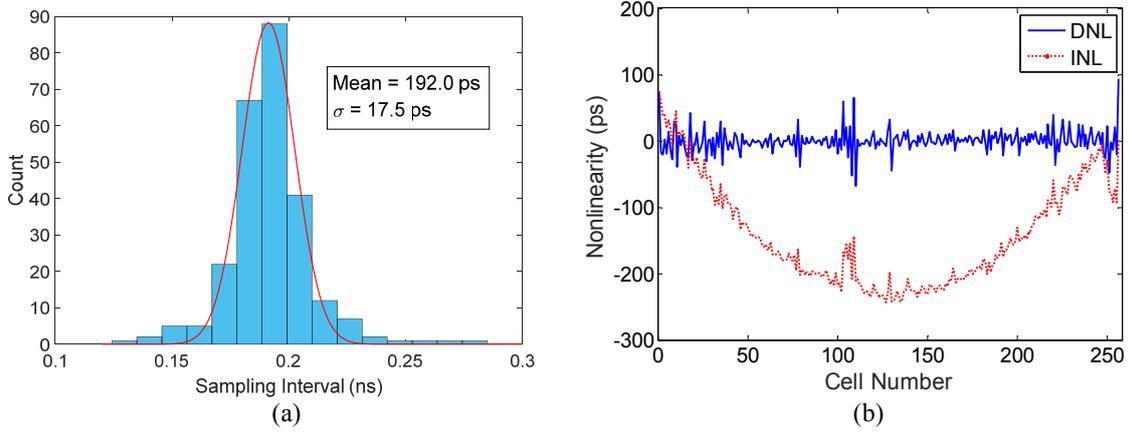

Fig. 7. Sampling interval calibration results of 256 cells. (a) Histogram of intervals. (b) DNL and INL of intervals.

By replacing the sampling intervals $\Delta t_i$ in Eq. (7) with the above calibration results, we recalculated the time difference $\Delta t_{ab}$ in Fig. 3, which should be equal to the period of the input sine wave. The calculated period values are plotted in Fig. 8 (b). For comparison, Fig. 8 (a) shows the calculated results without calibration. In Fig. 8 (a) and (b), the cell number in the abscissa is defined as the left of the two cells surrounding the first zero-crossing points in Fig. 3. For instance, the $\Delta t_{ab}$ obtained from Fig. 3 is represented by Cell No.172. Through comparison between Fig. 8 (a) and (b), the correction effect can be clearly observed.

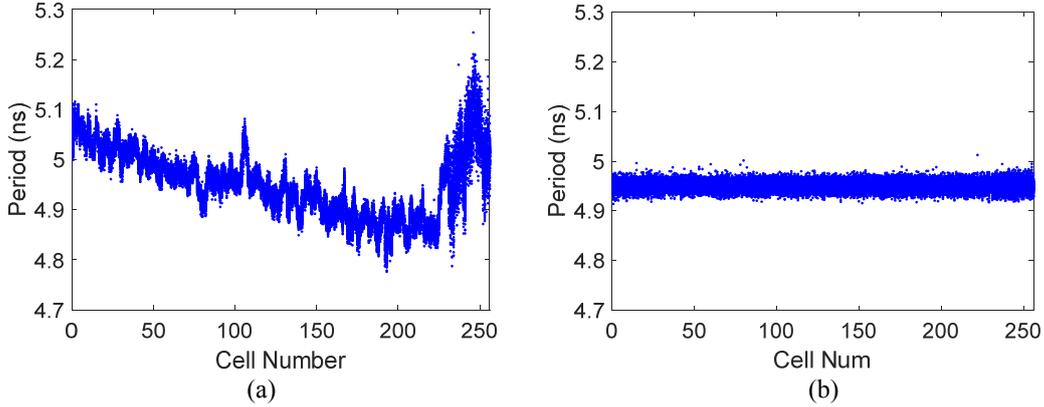

Fig. 8. Effect of the new calibration method when used to calculate the period of a 202 MHz sine wave. (a) Without calibration. (b) With calibration.

B. *Time measurement performance test results*

In the time measurement performance test, the Keysight 81160A signal source was used to generate a fast pulse which is further split to two pulses and sent to two sampling channels of the SCA ASIC. Through analyzing the waveforms digitized by these two channels, we calculated the time difference between these two pulses, as shown in Fig. 9 (a). We applied a polynomial functional fitting to the leading edge of each pulse, and a global threshold of 300 mV was used to obtain the time information. With a big amount of tests, the RMS value of the time difference test results can be calculated statistically. Considering that the time results of the two SCA channels are not interrelated, the time resolution of one single channel can be obtained by dividing the above RMS value by $1/\sqrt{2}$. Fig. 9 (b) shows the histogram of the time difference test results after correction with our method applied. The mean value is 3.93 ns, and the corresponding time resolution (of one single channel) is around 5.10 ps RMS.

We changed the time difference between the two input pulses from 0 ns to 20 ns and conducted a series of tests. The linearity of measured time difference is shown in Fig. 10. Fig. 10 (a) shows the measured time difference versus the known value, and Fig. 10 (b) shows the calculated INL. The INL is within 0.1% of the range 0-20 ns.

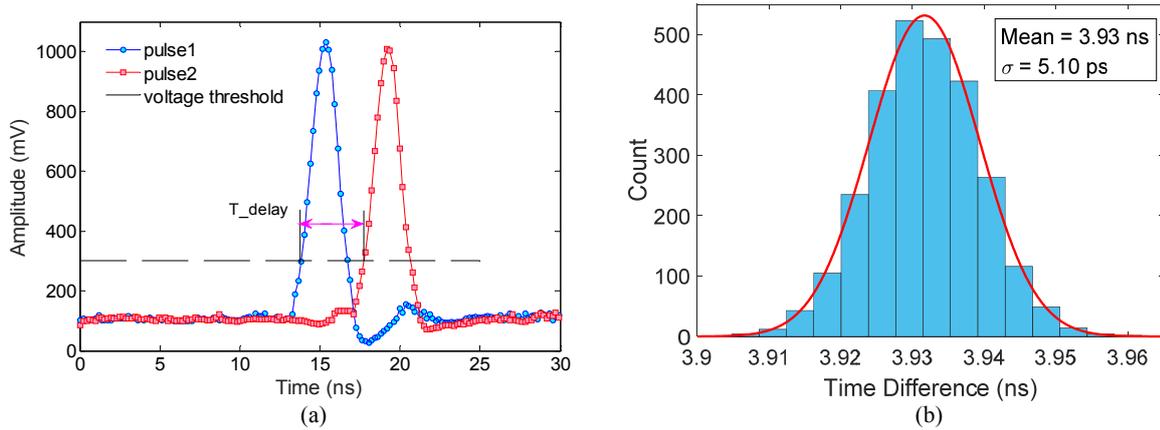

Fig. 9. (a) Waveforms recorded by two sampling channels. (b) Histogram of the time difference measurement results.

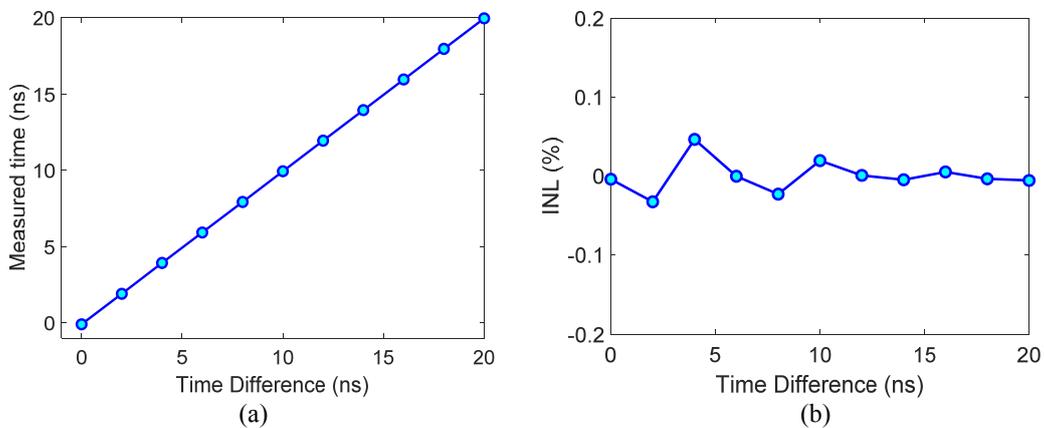

Fig. 10. Linearity of time measurement. (a) Measured time difference versus the known value. (b) INL.

Fig. 11 (a) shows the time resolution over an input time difference range from 0 to 20 ns. It indicates that after correction with the new time calibration method applied, the time resolution varies from 4 ps to 7 ps (RMS), all better than 10 ps. Furthermore, a comparison was made with other time calibration methods, and the results are shown in Fig. 11 (b). The time resolution using our method is the best and slightly better than the result using the global calibration method in Ref. [18], while our method features a simplified calibration process. The result using the zero-crossing method is even worse than that without calibration, and this phenomenon is probably due to the significant amplitude attenuation shown in Fig. 6. In Fig. 11 (a) and (b), it can also be observed that time resolution deteriorates when the time difference increases, and this is because that more samples are involved to calculate the time information and the residual errors of these sampling intervals accumulate, thus degrading the resolution.

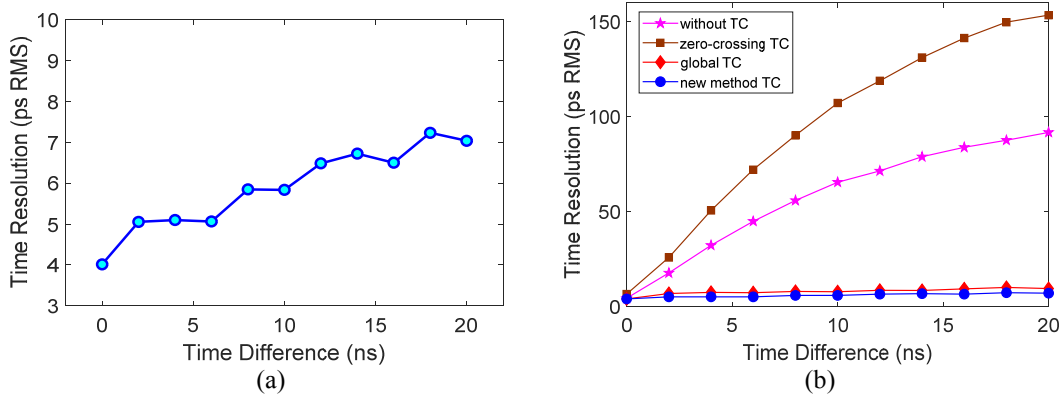

Fig. 11. Time resolution as the time difference varies from 0 to 20 ns. (a) Result of our new method. (b) Result comparison among different methods.

For high precision time measurement based on SCA ASICs, the time resolution is determined also by some other factors besides the time calibration, as described in [18]:

$$\Delta t = \frac{\Delta u}{U} \cdot \frac{1}{\sqrt{3 f_s \cdot f_{3dB}}} \qquad (11)$$

where $\Delta u$ is the voltage noise, $U$ is the signal amplitude, $f_s$ is the sampling speed and $f_{3dB}$ is the analog bandwidth of SCA. The sampling speed may be limited by the CMOS process or specific application, and the time resolution can be further improved by optimizing the voltage noise and analog bandwidth in the design.

## V. CONCLUSION

A new time calibration method for SCA-based waveform sampling ASICs is reported. This method can calibrate the sampling intervals by solving overdetermined linear equations. Furthermore, this method is almost not affected by the amplitude attenuation along the input bus caused by parasitic resistance and capacitance. By applying this method, the time resolution is improved significantly, and we achieved a time resolution of 4-7 ps RMS over a time difference range of 0-20 ns. Through comparison with other methods, this new method features the best performance and a simplified calibration process. The proposed time calibration method is expected to be applied to other waveform sampling ASICs based on the SCA architecture.

## ACKNOWLEDGEMENTS

The authors would like to thank Wei Wei in Institute of High Energy, CAS for his constant help in our ASIC design. We also appreciate the discussion during the previous SCA design work with Wei Wei, Zhi Deng of Department of Engineering Physics in Tsinghua University and others, which enhanced our understanding about SCA. We thank Fukun Tang of Enrico Fermi Institute in University of Chicago for his invaluable guidance in the previous SCA design work.

This work was supported in part by Key Research Program of Frontier Sciences, CAS under Grant QYZDB-SSW-SLH002, in part by Science and Technological Fund of Anhui Province for Outstanding Youth under Grant 1708085J07, in party by the Knowledge Innovation Program of the Chinese Academy of Sciences under Grant KJCX2-YW-N27, in part by the Fundamental Research Funds for the Central Universities under Grant WK2360000001, and in part by the CAS Center for Excellence in Particle Physics (CCEPP).